\documentstyle[aps]{revtex}
\begin{document}
\draft
\title{Explicit expressions of spin wave functions}
\author{Jie-Jie Zhu}
\address{
CCAST(World Lab), P. O. Box 8730,
Beijing, 100080, P. R. China\\
and
Center for Fundamental Physics, USTC,
Hefei, Anhui, 230026, P. R. China\footnote{Permanent adress.}
}
\date{\today}
\maketitle
\begin{abstract}
We derive the explicit expressions of the canonical and helicity wave
functions for massive particles with arbitrary spins. Properties of
these wave functions are discussed.
\end{abstract}

\section{Introduction}
To describe particles with high spins in amplitude analysis,
one needs to construct the explicit expressions of wave functions.
Detailed properties of the amplitudes are needed in tensor
analysis~\cite{ChungSF71,ChungTA93,FilippiniTA95,ZhuPHD,Zhua23pi} to give
independent invariant amplitudes free of kinematics singularities
and zeros~\cite{HaraKS64,WangKS66,StappKS67}.
We will give the explicit expressions of the canonical and
helicity wave functions for massive particles with
arbitrary spins in this paper. These wave functions satisfy
Rarita-Schwinger conditions~\cite{RaritaWF41}.

We will discuss quantum states in section~\ref{sec:qs}.
Spin wave functions are given in section~\ref{sec:iwf} and
section~\ref{sec:hwf}.

\section{Quantum States}\label{sec:qs}
Let $L(p)$ be a Lorentz transformation that satisfies
    \begin{equation}\label{eq:lrtz}
    p^\mu={L^\mu}_\nu(p)k^\nu.
    \end{equation}
For massive particles one can choose the standard momentum to be
$(k^\mu)=(w;\ \roarrow{0})$. $w$ is the mass of the particle. The
space-time metric is taken as
$(g^{\mu\nu})=diag\{1,-1,-1,-1\}.$
Now we can define one particle states as~\cite{WeinbergQFT1}
    \begin{equation}\label{eq:qsdef}
    |p,\sigma\rangle =U(L(p))|k,\sigma\rangle
    \end{equation}
with $U(L(p))$ the unitary representation of $L(p)$ in Hilbert
space. The one particle states satisfy
    \begin{equation}
    \hat{p}^\mu|p,\sigma\rangle=p^\mu|p,\sigma\rangle .
    \end{equation}

We choose the othonormal condition to be
    \begin{equation}
    \langle p^{'}, \sigma^{'} | p, \sigma \rangle =
    (2 \pi)^3(2 p^0) \delta(\roarrow{p}^{'} - \roarrow{p})
    \delta_{\sigma^{'}\sigma}.
    \end{equation}
Under Lorentz transformations,
    \begin{equation}
    U(\Lambda)| p,\sigma \rangle = \sum\limits_{\sigma^{'}}
      D_{\sigma^{'}\sigma}(W(\Lambda,p))|
       \Lambda p, \sigma^{'} \rangle;
    \end{equation}
where
    \begin{equation}
    D_{\sigma^{'}\sigma}(W(\Lambda,p))
        \equiv L^{-1}(\Lambda p)\Lambda L(p)
    \end{equation}
is the Wigner rotation~\cite{WignerLG39} and
$\{D_{\sigma^{'}\sigma}\}$ furnishes a representation of $SO(3).$
We also use the notation
$|\roarrow{p},j,m\rangle\equiv |p,j,m\rangle$.

There are infinite ways to define the Lorentz transformation that
satisfies equation~(\ref{eq:lrtz}). Canonical state and helicity
state are the two types mostly used.

If one define the Lorentz transformation in equation~(\ref{eq:qsdef})
to be pure Lorentz boost
    \begin{equation}\label{eq:qswr}
    \begin{array}{rcl}
    L(p)&=&L(\roarrow{p})\\
        &\equiv&R(\varphi,\theta,0)L_{z}
            (|\roarrow{p}|)R^{-1}(\varphi,\theta,0),
    \end{array}
    \end{equation}
the canonical state is obtained. Here $R(\varphi,\theta,0)$ is the
rotation that takes $z$-axis to the direction of $\roarrow{p}$ with
spherical angles $(\theta,\varphi)$, and the boost
$L_{z}(|\roarrow{p}|)$ takes the four-momentum
$(k^\mu)=(w;\ \roarrow{0})$ to
$\left( \sqrt{w^2+\roarrow{p}^2};\ 0,\ 0,\ |\roarrow{p}|\right)$.
For a particle of spin-$j$,$\sigma\sim(j,m)$.
It can be shown that for the canonical states,
equation~(\ref{eq:qswr}) become
    \begin{equation}
    U(\Lambda)|\roarrow{p},j,m\rangle=\sum\limits_{m^{'}}
      D_{m^{'} m}^j(L^{-1}(\roarrow{\Lambda}p)\Lambda L(\roarrow{p}))
      |\roarrow{\Lambda}p,j,m^{'}\rangle.
    \end{equation}
$ D_{m^{'} m}^j $ is the ordinary $D$-function. Especially, under
rotation $R$,
    \begin{equation}
    U(R)|\roarrow{p},j,m\rangle=\sum\limits_{m^{'}}
      D_{m^{'} m}^j(R)
      |\roarrow{R}p,j,m^{'}\rangle.
    \end{equation}

Defining the Lorentz transformation in another way will leads to
helicity states~\cite{JacobH59}:
    \begin{equation}
    \begin{array}{rcl}
    L(p)&=&L(\roarrow{p})R^{-1}(\varphi,\theta,0)\\
        &\equiv&R(\varphi,\theta,0)L_{z}(|\roarrow{p}|).
    \end{array}
    \end{equation}
We have
    \begin{equation}
    U(\Lambda)|\roarrow{p},j,\lambda\rangle=\sum\limits_{\lambda^{'}}
        D_{\lambda^{'}\lambda}^j
            (L^{-1}(\roarrow{\Lambda}p)\Lambda L(\roarrow{p}))
    |\roarrow{\Lambda}p,j,\lambda^{'}\rangle
    \end{equation}
and
    \begin{equation}
    U(R)|\roarrow{p},j,\lambda\rangle=
      |\roarrow{R}p,j,\lambda\rangle.
    \end{equation}

The two types of definitions are related to each other by
    \begin{equation}
    |\roarrow{p},j,\lambda\rangle_{helicity}
    =\sum\limits_m D^j_{m\lambda}(\varphi,\theta,0)
        |\roarrow{p},j,m\rangle_{canonical}.
    \end{equation}

We see that the definition of state depends on the choice of Lorentz
transformation in equation~(\ref{eq:lrtz}). There {\em is} a
definition called spinor state~\cite{JoosLG62}, which is different
from that of equation~(\ref{eq:qsdef})
and does not depend on the specific choice of Lorentz transformation;
but it makes things more complex and is seldom used.

Now we write quantum states in terms of  creation and
annihilation operators:
    \begin{equation}
    |\roarrow{p},\sigma\rangle=
      \sqrt{(2 \pi)^3 2 p^0}a^\dagger(\roarrow{p},\sigma)|0\rangle,
    \end{equation}
with $|0\rangle$ the vacuum state.
Quatum fields are given by~\cite{WeinbergQFT1}
\begin{eqnarray}
\psi_l^{(+)}=\int\frac{d^3 p}{\sqrt{\left(2\pi\right)^3 2 p^0}}
  \sum\limits_\sigma U_l(\roarrow{p},\sigma)a(\roarrow{p},\sigma)
  e^{-ip\cdot x},\\
\psi_l^{(-)}=\int\frac{d^3 p}{\sqrt{\left(2\pi\right)^3 2 p^0}}
  \sum\limits_\sigma V_l(\roarrow{p},\sigma)a^\dagger
    (\roarrow{p},\sigma)e^{ip\cdot x},\\
U(\Lambda,a)\psi_l^{(\pm)}U^{-1}(\Lambda,a)=
  \sum\limits_{l^{'}}G_{l l^{'}}(\Lambda^{-1})
  \psi_{l^{'}}^{(\pm)}(\Lambda x +a).
\end{eqnarray}
The coefficient functions, $U_l$ and $V_l$, are wave functions in
momentum space. $a^\mu$ are parameters for translation.
$\{G_{l l^{'}}\}$ furnishes a representation of the Lorentz group.
One finds that wave functions satisfy~\cite{WeinbergQFT1}
\begin{eqnarray}
\sum\limits_{l^{'}}G_{l l^{'}}(\Lambda)U_{l^{'}}(\roarrow{p},\sigma)
=\sum\limits_{\sigma^{'}}D_{\sigma^{'}\sigma}(W(\Lambda,p))
    U_l(\roarrow{\Lambda}p,\sigma^{'}),\\
\sum\limits_{l^{'}}G_{l l^{'}}(\Lambda)V_{l^{'}}(\roarrow{p},\sigma)
=\sum\limits_{\sigma^{'}}D^*_{\sigma^{'}\sigma}(W(\Lambda,p))
    V_l(\roarrow{\Lambda}p,\sigma^{'});
\end{eqnarray}
so we can define wave functions as
\begin{eqnarray}
U_l(\roarrow{p},\sigma) & = &
  \sum\limits_{l^{'}}G_{l l^{'}}
    (L(\roarrow{p}))U_{l^{'}}(\roarrow{k},\sigma),\\
V_l(\roarrow{p},\sigma) & = &
  \sum\limits_{l^{'}}G_{l l^{'}}
    (L(\roarrow{p}))V_{l^{'}}(\roarrow{k},\sigma).
\end{eqnarray}
For massive particles, $\roarrow{k}=\roarrow{0}$.

\section{Wave Functions for IntegralL Spin Particles}\label{sec:iwf}
If the index $l$ in previous section is chosen as Lorentz indexes,
one arrived at vector fields:
    \begin{equation}
    \begin{array}{c}
    {G(\Lambda)^\mu}_\nu={\Lambda^\mu}_\nu,\\
    U^\mu(\roarrow{p},\sigma)=
        {L(p)^\mu}_\nu U^\nu(\roarrow{0},\sigma),\\
    V^\mu(\roarrow{p},\sigma)=
        {L(p)^\mu}_\nu V^\nu(\roarrow{0},\sigma).
    \end{array}
    \end{equation}

We use the following infinitesimal generators of the Lorentz group:
    \begin{equation}
    \begin{array}{ccc}
    \left({{J_1}^\mu}_\nu\right)= 
      \left(
      \begin{array}{cccc}
      0 & 0 & 0 & 0\\
      0 & 0 & 0 & 0\\
      0 & 0 & 0 & -i\\
      0 & 0 & i & 0
      \end{array}
      \right), &
    \left({{J_2}^\mu}_\nu\right)= 
      \left(
      \begin{array}{cccc}
      0 & 0 & 0 & 0\\
      0 & 0 & 0 & i\\
      0 & 0 & 0 & 0\\
      0 & -i & 0 & 0
      \end{array}
      \right), &
    \left({{J_3}^\mu}_\nu\right)= 
      \left(
      \begin{array}{cccc}
      0 & 0 & 0 & 0\\
      0 & 0 & -i & 0\\
      0 & i & 0 & 0\\
      0 & 0 & 0 & 0
      \end{array}
      \right), \\
    \left({{K_1}^\mu}_\nu\right)= 
      \left(
      \begin{array}{cccc}
      0 & i & 0 & 0\\
      i & 0 & 0 & 0\\
      0 & 0 & 0 & 0\\
      0 & 0 & 0 & 0
      \end{array}
      \right), &
    \left({{K_2}^\mu}_\nu\right)= 
      \left(
      \begin{array}{cccc}
      0 & 0 & i & 0\\
      0 & 0 & 0 & 0\\
      i & 0 & 0 & 0\\
      0 & 0 & 0 & 0
      \end{array}
      \right), &
    \left({{K_3}^\mu}_\nu\right)= 
      \left(
      \begin{array}{cccc}
      0 & 0 & 0 & i\\
      0 & 0 & 0 & 0\\
      0 & 0 & 0 & 0\\
      i & 0 & 0 & 0
      \end{array}
      \right);
    \end{array}
    \end{equation}
and get the explicit expressions of canonical wave functions ($E$
is the energy of the particle)
    \begin{equation}
    \begin{array}{rcl}
    \left(e_c^\mu(\roarrow{p},0)\right) & = &
          \left(
          \begin{array}{c}
            \frac{|\roarrow{p}|}{w}\cos\theta\\
            \frac{1}{2}\left(\frac{E}{w}-1\right)
                \sin 2\theta\cos\varphi\\
            \frac{1}{2}\left(\frac{E}{w}-1\right)
                \sin 2\theta\sin\varphi\\
            \frac{1}{2}\left(\frac{E}{w}-1\right)(1+\cos 2\theta)+1
          \end{array}  
          \right),\\
    \left(e_c^\mu(\roarrow{p},\pm 1)\right) & = &
          \mp\frac{1}{\sqrt{2}}\left(
          \begin{array}{c}
            \frac{|\roarrow{p}|}{w}\sin\theta e^{\pm i\varphi}\\
            \left(\frac{E}{w}-1\right)\sin^2\theta\cos\varphi
                    e^{\pm i\varphi}+1\\
            \left(\frac{E}{w}-1\right)\sin^2\theta\sin\varphi
                    e^{\pm i\varphi}\pm 1\\
            \left(\frac{E}{w}-1\right)\cos\theta\sin\theta
                    e^{\pm i\varphi}
          \end{array}  
          \right);
    \end{array}
    \end{equation}
while helicity wave functions are
\begin{equation}
    \begin{array}{rcl}
    \left(e_h^\mu(\roarrow{p},0)\right) & = &
          \left(
          \begin{array}{c}
            \frac{|\roarrow{p}|}{w}\\
            \frac{E}{w}\sin \theta\cos\varphi\\
            \frac{E}{w}\sin \theta\sin\varphi\\
            \frac{E}{w}\cos \theta
          \end{array}  
          \right),\\
    \left(e_h^\mu(\roarrow{p},\pm 1)\right) & = &
          \frac{1}{\sqrt{2}}\left(
          \begin{array}{c}
            0\\
            \mp\cos\theta\cos\varphi +i\sin\varphi\\
            \mp\cos\theta\sin\varphi -i\cos\varphi\\
            \pm\sin\theta
          \end{array}  
          \right).
    \end{array}
\end{equation}
We have
    \begin{equation}
    U(\roarrow{p},\sigma)=V^{*}(\roarrow{p},\sigma)
      =e(\roarrow{p},\sigma).
    \end{equation}

Wave functions for higher integral spins can be defined recurrently by
\begin{equation}
      e_{\mu_1\mu_2\cdots\mu_j}(\roarrow{p},j,\sigma)
    =\sum\limits_{\sigma^{'}_{j-1},\sigma_j}
      (j-1,\sigma^{'}_{j-1};1,\sigma_j|j,\sigma)
      e_{\mu_1\mu_2\cdots\mu_{j-1}}(\roarrow{p},j-1,\sigma_{j-1}^{'})
      e_{\mu_j}(\roarrow{p},\sigma_j).
\end{equation}
Using the C-G coefficient relation
\begin{equation}
\begin{array}{rl}
 & \sum\limits_{\sigma^{'}_3, \sigma^{'}_4, \cdots, \sigma^{'}_n}
  (j_1,\sigma_1;j_2,\sigma_2|j_1+j_2,\sigma^{'}_3)
  (j_1+j_2,\sigma^{'}_3;k_3,\sigma_3|j_1+j_2+j_3,\sigma^{'}_4)\cdots\\
 & \ \ \ \times
  (j_1+j_2+\cdots+j_{n-1},\sigma^{'}_n;
        j_n,\sigma_n|j_1+j_2+\cdots+j_n,
        \sigma^{'}_n+\sigma_n)\\
=&\left[\prod\limits_{i=1}^{n}
    \frac{(2j_i)!}{(j_i+\sigma_i)! (j_i-\sigma_i)!}\right]^\frac{1}{2}
   \left\{\frac{\left[\sum\limits^n_{i=1}(j_i+\sigma_i)\right]!
            \left[\sum\limits^n_{i=1}(j_i-\sigma_i)\right]!}
           {\left(2\sum\limits_{i=1}^n j_i\right)!}
   \right\}^\frac{1}{2},
\end{array}
\end{equation}
we find
\begin{equation}
\begin{array}{rl}
 &  e_{\mu_1\mu_2\cdots\mu_j}(\roarrow{p},j,\sigma) \\
=& \sum\limits_{\sigma_1, \sigma_2,\cdots, \sigma_j}
   \left\{ \frac{2^j(j+\sigma)!(j-\sigma)!}
      {(2j)!\prod\limits^j_{i=1}[(1+\sigma_i)!(1-\sigma_i)!]}
   \right\}^{\frac{1}{2}}
    \delta_{\sigma_1+\sigma_2+\cdots+\sigma_j,\sigma}
  e_{\mu_1}(\roarrow{p},\sigma_1)
                e_{\mu_2}(\roarrow{p},\sigma_2)\cdots
     e_{\mu_j}(\roarrow{p},\sigma_j).
\end{array}
\end{equation}

It is easy to show
\begin{equation}
  {\Lambda^{\mu_1}}_{\nu_1}{\Lambda^{\mu_2}}_{\nu_2}\cdots
     {\Lambda^{\mu_j}}_{\nu_j}
     e^{\nu_1\nu_2\cdots\nu_j}(\roarrow{p},j,\sigma) =
  \sum\limits_{\sigma^{'}}
    D^j_{\sigma^{'}\sigma}(W(\Lambda,\roarrow{p}))
  e^{\mu_1\mu_2\cdots\mu_j}(\roarrow{\Lambda}p,j,\sigma^{'}).
\end{equation}

$e_{\mu_1\mu_2\cdots\mu_j}(\roarrow{p},j,\sigma)$ satisfies all of the
Rarita-Schwinger conditions: space-like, symmetric and traceless.

\section{Wave Functions for
Half-integral Spin Particles}\label{sec:hwf}
The convention of $\gamma$-matrices used here follows that of 
Bjorken and Drell~\cite{BjorkenQF}, so the generators of the Lorentz 
group are
\begin{equation}
\begin{array}{cc}
\roarrow{J}=\frac{1}{2}
   \left(\begin{array}{cc}
    \roarrow{\tau} & 0 \\ 0 & \roarrow{\tau}
   \end{array}\right), &
\roarrow{K}=\frac{i}{2}
   \left(\begin{array}{cc}
   0 & \roarrow{\tau} \\ \roarrow{\tau} & 0
   \end{array}\right);
\end{array}
\end{equation}
with $\roarrow{\tau}$ Pauli matrixes.

The spin-$\frac{1}{2}$ canonical wave functions are (
here $\alpha=\ln\left((E+|\roarrow{p}|)/w\right)$)
\begin{equation}
\begin{array}{cc}
U_c(\roarrow{p},\frac{1}{2})=
  \left(\begin{array}{c}
    \cosh\frac{\alpha}{2}\\ 0\\ \cos\theta\sinh\frac{\alpha}{2}\\
    \sin\theta e^{i\varphi}\sinh\frac{\alpha}{2}
  \end{array}\right),&
U_c(\roarrow{p},-\frac{1}{2})=
  \left(\begin{array}{c}
    0\\ \cosh\frac{\alpha}{2}\\ 
    \sin\theta e^{-i\varphi}\sinh\frac{\alpha}{2}\\
    -\cos\theta\sinh\frac{\alpha}{2}
  \end{array}\right);\\
V_c(\roarrow{p},\frac{1}{2})=
  \left(\begin{array}{c}
  \sin\theta e^{-i\varphi}\sinh\frac{\alpha}{2}\\
  -\cos\theta\sinh\frac{\alpha}{2}\\
  0\\ \cosh\frac{\alpha}{2}
  \end{array}\right), &
V_c(\roarrow{p},-\frac{1}{2})=
  \left(\begin{array}{c}
  -\cos\theta\sinh\frac{\alpha}{2}\\
  -\sin\theta e^{i\varphi}\sinh\frac{\alpha}{2}\\
  -\cosh\frac{\alpha}{2}\\ 0
  \end{array}\right);
\end{array}
\end{equation}
and the helicity wave functions are
\begin{equation}
\begin{array}{ll}
U_h(\roarrow{p},\frac{1}{2})=
  \left(\begin{array}{c}
  \cos\frac{\theta}{2}e^{-i\frac{\varphi}{2}}\cosh\frac{\alpha}{2}\\
  \sin\frac{\theta}{2}e^{i\frac{\varphi}{2}}\cosh\frac{\alpha}{2}\\
  \cos\frac{\theta}{2}e^{-i\frac{\varphi}{2}}\sinh\frac{\alpha}{2}\\
  \sin\frac{\theta}{2}e^{i\frac{\varphi}{2}}\sinh\frac{\alpha}{2}
  \end{array}\right),&
U_h(\roarrow{p},-\frac{1}{2})=
  \left(\begin{array}{c}
  -\sin\frac{\theta}{2}e^{-i\frac{\varphi}{2}}\cosh\frac{\alpha}{2}\\
  \cos\frac{\theta}{2}e^{i\frac{\varphi}{2}}\cosh\frac{\alpha}{2}\\
  \sin\frac{\theta}{2}e^{-i\frac{\varphi}{2}}\sinh\frac{\alpha}{2}\\
  -\cos\frac{\theta}{2}e^{i\frac{\varphi}{2}}\sinh\frac{\alpha}{2}
  \end{array}\right);\\
V_h(\roarrow{p},\frac{1}{2})=
  \left(\begin{array}{c}
  \sin\frac{\theta}{2}e^{-i\frac{\varphi}{2}}\sinh\frac{\alpha}{2}\\
  -\cos\frac{\theta}{2}e^{i\frac{\varphi}{2}}\sinh\frac{\alpha}{2}\\
  -\sin\frac{\theta}{2}e^{-i\frac{\varphi}{2}}\cosh\frac{\alpha}{2}\\
  \cos\frac{\theta}{2}e^{i\frac{\varphi}{2}}\cosh\frac{\alpha}{2}
  \end{array}\right),&
V_h(\roarrow{p},-\frac{1}{2})=
  \left(\begin{array}{c}
  -\cos\frac{\theta}{2}e^{-i\frac{\varphi}{2}}\sinh\frac{\alpha}{2}\\
  -\sin\frac{\theta}{2}e^{i\frac{\varphi}{2}}\sinh\frac{\alpha}{2}\\
  -\cos\frac{\theta}{2}e^{-i\frac{\varphi}{2}}\cosh\frac{\alpha}{2}\\
  -\sin\frac{\theta}{2}e^{i\frac{\varphi}{2}}\cosh\frac{\alpha}{2}
\end{array}\right).
\end{array}
\end{equation}

Spin-$n+\frac{1}{2}$ wave functions read
\begin{equation}
\begin{array}{rl}
&U_{\mu_1\mu_2\cdots\mu_n}(\roarrow{p},n+\frac{1}{2},\sigma)\\ = &
  \sum\limits_{\sigma_1,\sigma_2,\cdots, \sigma_{n+1}}
  \left\{\frac{2^n(n+\frac{1}{2}+\sigma)!(n+\frac{1}{2}-\sigma)!}
      {(2n+1)!\prod\limits^n_{i=1}[(1+\sigma_i)!(1-\sigma_i)!]}
  \right\}^\frac{1}{2}
  \delta_{\sigma_1+\sigma_2+\cdots+\sigma_{n+1},\sigma}\\
& \ \ \ \times e_{\mu_1}(\roarrow{p},\sigma_1)
                e_{\mu_2}(\roarrow{p},\sigma_2)
  \cdots e_{\mu_n}(\roarrow{p},\sigma_n)U(\roarrow{p},\sigma_{n+1});
\end{array}
\end{equation}
\begin{equation}
\begin{array}{rl}
&V_{\mu_1\mu_2\cdots\mu_n}(\roarrow{p},n+\frac{1}{2},\sigma)\\ = &
  \sum\limits_{\sigma_1,\sigma_2,\cdots, \sigma_{n+1}}
  \left\{\frac{2^n(n+\frac{1}{2}+\sigma)! (n+\frac{1}{2}-\sigma)!}
      {(2n+1)!\prod\limits^n_{i=1}[(1+\sigma_i)!(1-\sigma_i)!]}
  \right\}^\frac{1}{2}
  \delta_{\sigma_1+\sigma_2+\cdots+\sigma_{n+1},\sigma}\\
& \ \ \ \times e^*_{\mu_1}(\roarrow{p},\sigma_1)
               e^*_{\mu_2}(\roarrow{p},\sigma_2)
  \cdots e^*_{\mu_n}(\roarrow{p},\sigma_n)V(\roarrow{p},\sigma_{n+1}).
\end{array}
\end{equation}
They satisfy Dirac equations and Rarita-Schwinger
conditions~\cite{RaritaWF41}; especially
\begin{equation}
\gamma^{\mu_k}U_{\mu_1\mu_2\cdots\mu_k\cdots\mu_n}
   (\roarrow{p},n+\frac{1}{2},\sigma)=0,
\end{equation}
\begin{equation}
\gamma^{\mu_k}V_{\mu_1\mu_2\cdots\mu_k\cdots\mu_n}
   (\roarrow{p},n+\frac{1}{2},\sigma)=0.
\end{equation}


\begin{references}
\bibitem{ChungSF71}S. U. Chung,
    {\em Spin Formalisms}, CERN 71-8 (1971).
\bibitem{ChungTA93}S. U. Chung,
    Phys. Rev. D {\bf 48}, 1225 (1993).
\bibitem{FilippiniTA95}V. Filippini, A. Fontana and A. Rotondi,
    Phys. Rev. D {\bf 51}, 2247 (1995).
\bibitem{ZhuPHD}Jie-Jie Zhu,  {\em Ph.D thesis},
    (University of Science and Techonology of China, 1997), unpublished.
\bibitem{Zhua23pi}Jie-Jie Zhu, 
    Phys. Rev. D {\bf 57}, 5468 (1998). 
\bibitem{HaraKS64}Y. Hara,
    Phys. Rev. {\bf 136}, B507 (1964).
\bibitem{WangKS66}L. L. C. Wang,
    Phys. Rev. {\bf 142}, 1187 (1966).
\bibitem{StappKS67}H. P. Stapp,
    Phys. Rev. {\bf 160}, 1251 (1967).
\bibitem{RaritaWF41}W. Rarita and J. Schwinger,
    Phys. Rev. {\bf 60}, 61 (1941).
\bibitem{WeinbergQFT1}Steven Weinberg,
{\em The Quantum Theory of Fields, Vol. I, Foundations},
(Cambridge University Press, Cambridge, 1995).
\bibitem{WignerLG39}E. P. Wigner,
    Ann. Math. {\bf 40}, 149 (1939).
\bibitem{JacobH59}M. Jacob and G. C. Wick,
    Ann. Phys. {\bf 7}, 404 (1959).
\bibitem{JoosLG62}H. Joos,
    Fortschr. Physik {\bf 10}, 65 (1962).
\bibitem{BjorkenQF}J. D. Bjorken and S. D. Drell,
    {\em Relativistic Quatum Fields},
    (MacGraw-Hill, New York, 1965).
\end{references}
\end{document}